\documentclass[a4paper]{jpconf}
\usepackage{graphicx}
\usepackage{subfig}
\usepackage{hyperref}

\begin{document}

\title{The Auger Engineering Radio Array and multi-hybrid cosmic ray detection}
\author{E M Holt\textsuperscript{1} for the Pierre Auger Collaboration\textsuperscript{2}}
\address{\textsuperscript{1} Institute of Nuclear Physics, Karlsruhe Institute of Technology, Karlsruhe, Germany}
\address{\textsuperscript{2} Observatorio Pierre Auger, Av. San Mart\'{i}n Norte 304, 5613 Malarg\"{u}e, Argentina}
\ead{ewa.holt@kit.edu}
\address{Full author list: \url{http://www.auger.org/archive/authors_2015_09.html}}

\begin{abstract}
The Auger Engineering Radio Array (AERA) aims at the detection of air showers induced by high-energy cosmic rays.
As an extension of the Pierre Auger Observatory, it measures complementary information to the particle detectors, fluorescence telescopes and to the muon scintillators of the Auger Muons and Infill for the Ground Array (AMIGA).
AERA is sensitive to all fundamental parameters of an extensive air shower such as the arrival direction, energy and depth of shower maximum.
Since the radio emission is induced purely by the electromagnetic component of the shower, in combination with the AMIGA muon counters, AERA is perfect for separate measurements of the electrons and muons in the shower, if combined with a muon counting detector like AMIGA.
In addition to the depth of the shower maximum, the ratio of the electron and muon number serves as a measure of the primary particle mass.
\end{abstract}

\section{Introduction}
The Auger Engineering Radio Array (AERA) is a modern radio experiment for the detection of cosmic ray air showers. 
It is the radio extension of the Pierre Auger Observatory, located in the Province of Mendoza in Argentina.
With its area of 17\,km$^2$ AERA is the world's largest experiment in the field of cosmic ray radio detection.
The coincident measurement with the other low energy extensions of the Pierre Auger Observatory \cite{auger} enables the simultaneous measurement of a number of air shower properties.\par
The Pierre Auger Observatory is an experiment for ultra-high-energy cosmic rays \cite{auger}.
It combines different detection techniques to obtain complementary information about extensive cosmic ray air showers.
1660 water-Cherenkov stations form the surface detector (SD) and are distributed with a spacing of 1500\,m over an area of 3000\,km$^2$.
They measure the particles of the showers at ground.
Extensive air showers induce radiation in the ultraviolet range due to excitation of atmospheric nitrogen by the shower particles.
On moonless clear nights, this radiation is detected by 27 fluorescence telescopes at four sites, overlooking the SD area and enabling the hybrid detection of showers.\par
Several enhancements, aiming at lower energies to 10$^{17}$\,eV, were installed in one part of the Observatory.
The Auger Muons and Infill for the Ground Array (AMIGA) \cite{AMIGA} covers an area of about 20\,km$^2$.
In AMIGA the spacing between the water-Cherenkov stations is reduced to 750\,m, yielding full efficiency for air showers with a primary energy down to $10^{17.5}$\,eV.
Buried muon scintillators, at 2.3\,m depth, accompany the water-Cherenkov stations for a better separation of electrons and muons in a shower.
Seven stations with muon detectors, forming the "Unitary Cell", have been taking data since 2013.
Three high-elevation air fluorescence telescopes (High Elevation Auger Telescope -- HEAT), observe low energy showers higher in the atmosphere.
AERA is located inside the region of the dense stations, enabling a combined detection of showers and cross calibration of the detectors.
In addition, AERA uses the particle detectors as an external trigger.\par
The radio emission of air showers is mainly induced by two mechanisms:
a) the geomagnetic effect due to deflection of the charged particles of the shower in the geomagnetic field \cite{geomagn}, and b) the Askaryan effect due to positron annihilation and ionization of atmospheric molecules by the shower particles, which cause an excess of negative charges in the shower front \cite{askaryan}.
Hence, the radio emission contains information on the development of the shower, in particular the position of the shower maximum X$_{\textrm{max}}$.
In first order, the radio emission is caused purely by the electromagnetic part of the shower.\par
Data taking is possible for almost 100\,\% of the time.
Hence, in comparison with the fluorescence detector (FD), an X$_{\textrm{max}}$ measurement is possible around the clock.
Furthermore, radio detection -- in contrast to particle detection -- becomes more efficient for more inclined showers due to a larger footprint of the emission at ground \cite{inclined}.

\section{The Auger Engineering Radio Array - detector description}
\begin{figure}
\subfloat[]{ \includegraphics[width=0.57\textwidth]{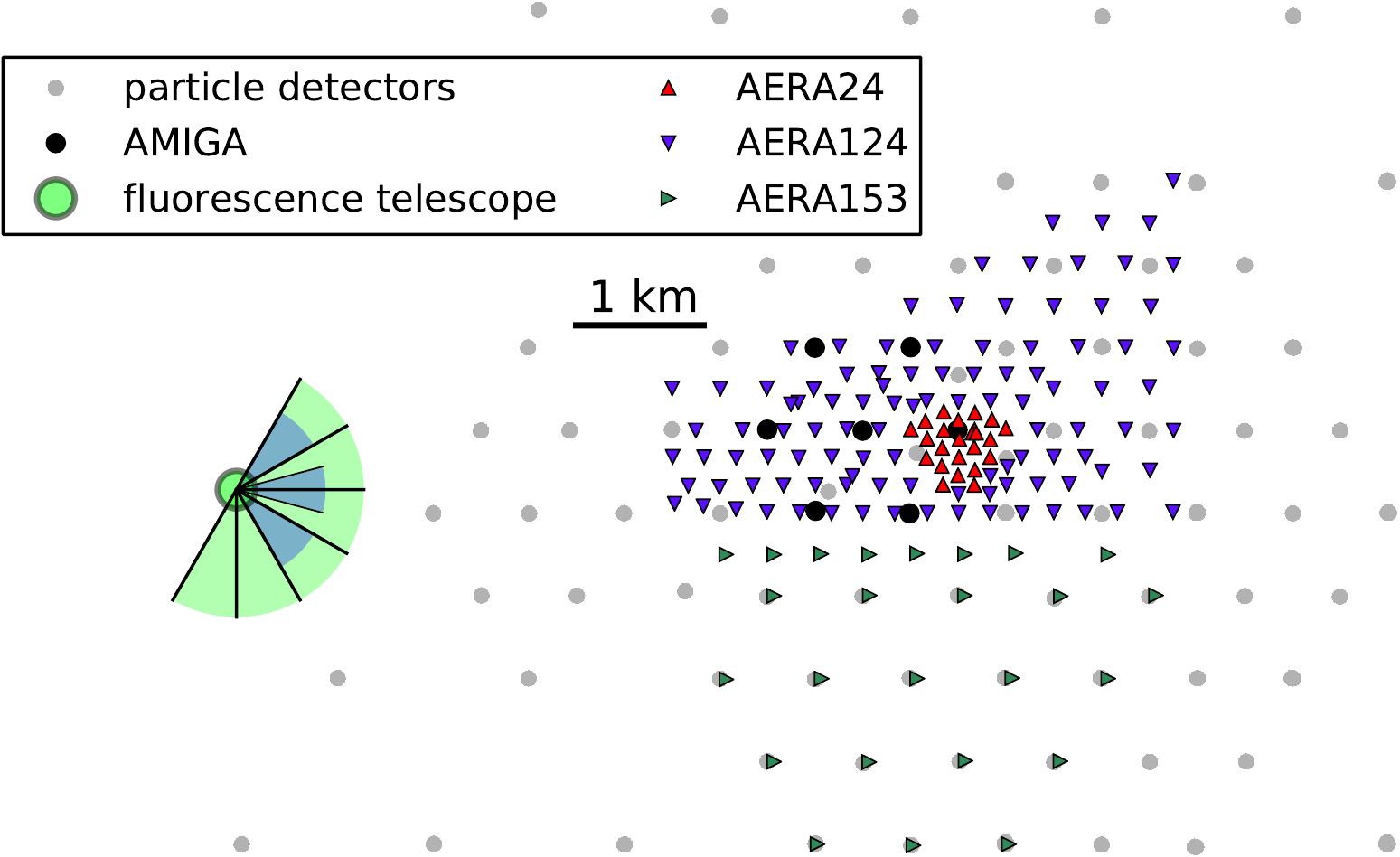}
\label{fig:AERAmap}
}
\hspace{0.02\textwidth}
\subfloat[]{\includegraphics[width=0.41\textwidth]{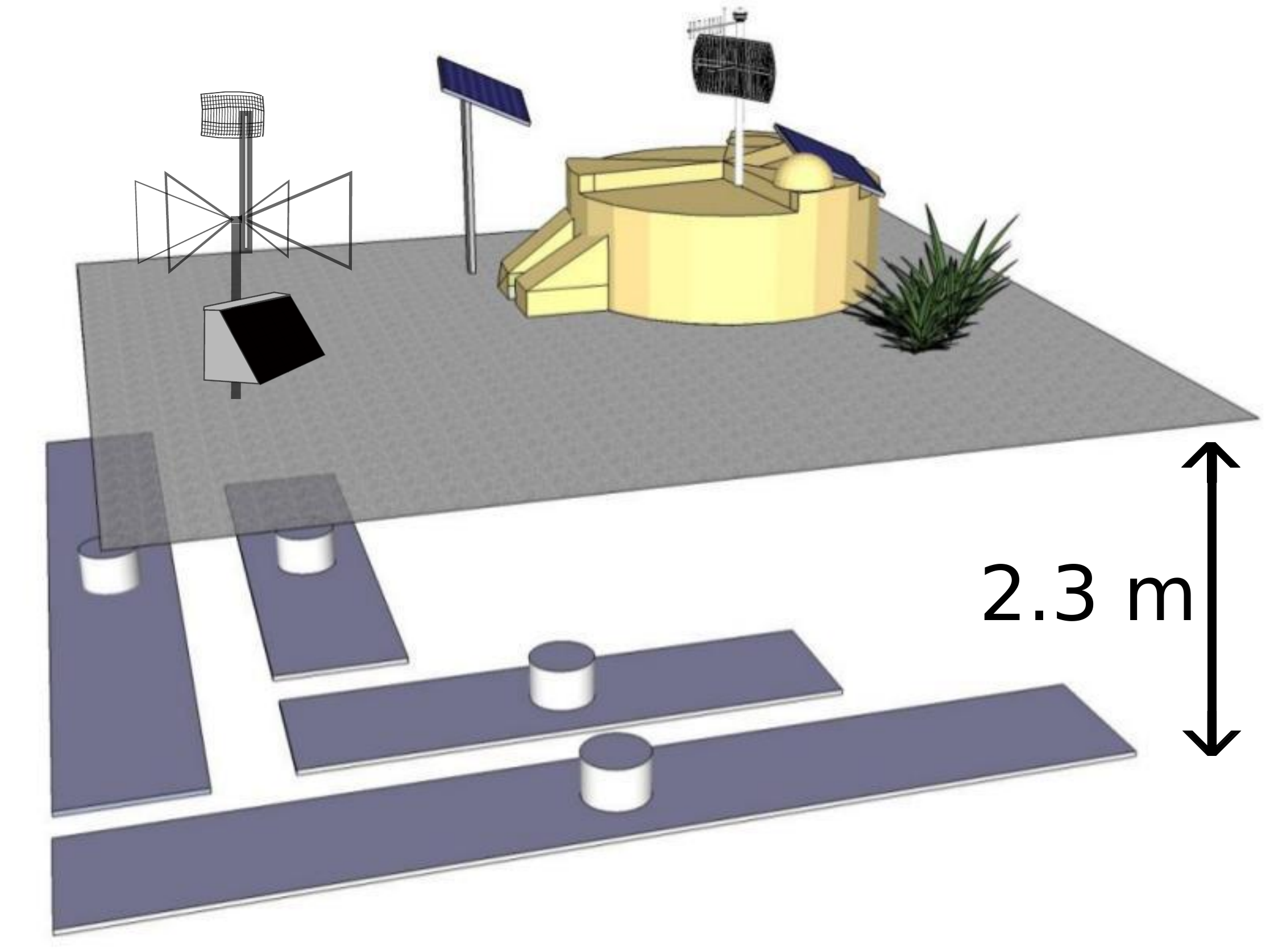}
\label{fig:detectors}
}
\caption{\protect\subref{fig:AERAmap} Layout of AERA {together} with the colocated water-Cherenkov stations, fluorescence telescopes and AMIGA Unitary Cell and \protect\subref{fig:detectors} sketch of an AERA butterfly antenna with a water-Cherenkov tank and buried muon scintillators of AMIGA.}
\end{figure}

AERA was built in three phases.
Starting in September 2010, AERA24 was deployed with 24 radio detection stations (RDS) featuring logarithmic periodic dipole antennas (LPDA) \cite{LPDA} with a spacing of 144\,m between the stations. 
AERA24 was used to investigate the radio emission itself and to develop techniques for radio detection of cosmic rays.
For the second phase, AERA124, another 100 RDS were installed in May 2013.
These RDS feature a different antenna type, the so called butterfly antenna \cite{butterfly}, and improved hardware.
They are distributed with spacings of 250\,m and 375\,m, and together with the first 24 antennas, cover an area of about 6\,km$^2$.
With this configuration AERA measures several thousand cosmic ray events per year from a primary energy of about 10$^{17}$\,eV up to the highest energies.
In March 2015, 25 additional butterfly antenna stations were installed on a grid with a 750\,m spacing, aiming mainly on the detection of horizontal air showers (\textgreater\,55$^{\circ}$ zenith angle).
With some additional prototype stations, AERA153 now consists of 153 RDS on an array of about 17\,km$^2$.
A map of AERA with the different phases and the other detectors of the Pierre Auger Observatory is shown in figure \ref{fig:AERAmap}.
An AERA butterfly antenna together with a water-Cherenkov station and buried muon scintillator of AMIGA is sketched in figure \ref{fig:detectors}. \par
All stations of AERA are equipped with a solar panel, battery and signal processing hardware.
They work completely autonomous in the field and send the collected data upon request to a central data facility via a WiFi link.
The antennas are aligned along the magnetic north-south and east-west direction.
They are triggered externally by the particle and fluorescence detectors and from internal triggers.
The signals are bandpass-filtered to the range of 30 -- 80\,MHz. 

\section{Results from AERA}
AERA was built for several different purposes: to improve the understanding of the radio emission mechanisms, for cosmic ray physics in the transition region between galactic and extragalactic cosmic rays, and to test the feasibility of a large-scale radio array for the highest energies. 

\subsection{Probing the theory of radio emission}
The different radio-emission mechanisms produce differently polarized radio emission -- linearly polarized from the geomagnetic effect and radial polarized towards the shower axis from the Askaryan effect.
By measuring the polarization of the radio emission, it is possible to study the contributions from the different effects.
Since the contribution from the geomagnetic effect depends on the angle to the geomagnetic field, this contribution is different for different detector sites.
Polarization measurements in AERA revealed a radial component with a mean contribution of 14\,\% \cite{polarization}, aside from the linear polarization.
This measurements agree well with the theory of the geomagnetic and Askaryan emission processes.

\subsection{Properties of the primary cosmic ray particle}
The main properties of a cosmic ray are its direction, energy and mass.
Radio measurements of AERA are sensitive to all of these properties.
The reconstruction of these properties is done by the Auger Offline software framework \cite{Offline}, which includes all detector systems for combined analysis.\par

{\it Arrival direction of the primary cosmic ray particle:}
The shower axis corresponds to the arrival direction of the primary particle.
This axis is reconstructed from the timing information of the radio pulses in the radio stations using a wavefront model for the radio emission.
For this, a plane wave is used as first-order approximation for the wavefront shape.
The reconstructed direction is in good agreement with the direction from the SD.\par

{\it Energy of the primary cosmic ray particle:}
The energy contained in the radio emission yields information about the primary cosmic ray energy.
To calculate the radiation energy, the measured electric-field strength of the 30 -- 80\,MHz radiation at the station positions is converted to the energy density.
We use a two-dimensional lateral distribution function (2D-LDF) \cite{2dldf} taking into account asymmetries due to the combined geomagnetic and Askaryan effect, to interpolate the energy density.
The integral over this 2D-LDF corresponds to the total radiation energy.
A calibration against the SD shows that the radiation energy is 15.9\,MeV for a cosmic ray energy of 1\,EeV \cite{energy}.
It scales quadratically with the cosmic ray energy because of the coherent character of the emission.
This radiation energy can therefore be used as an energy estimator.
In AERA, an energy resolution of 22\,\% for a dataset of AERA24 events, and of 17\,\% for a subset of events with high multiplicity ($\geq$ 5 radio stations) has been found.\par

{\it Shower maximum and mass composition:}
The depth in the atmosphere at which the number of secondary particles is maximum is called the shower maximum X$_{\textrm{max}}$.
It is strongly correlated to the mass of the primary particle.
The radio emission is mainly produced around and before this X$_{\textrm{max}}$ \cite{Scholten:Ludwig} and therefore carries information about it.
In AERA, there are different studies ongoing to reconstruct X$_{\textrm{max}}$ from the radio emission, e.g. using shape parameters of the hyperbolic radio wavefront \cite{Qader}, the width of the radio footprint in the shower plane \cite{johannesICRC}, or the slope of the frequency spectrum in single radio stations \cite{spectralSlope}.
The X$_{\textrm{max}}$ measurements by the FD are used for calibration and comparison of the results.

\section{Mass composition with multi-hybrid detection}

\begin{figure}
\includegraphics[width=\textwidth]{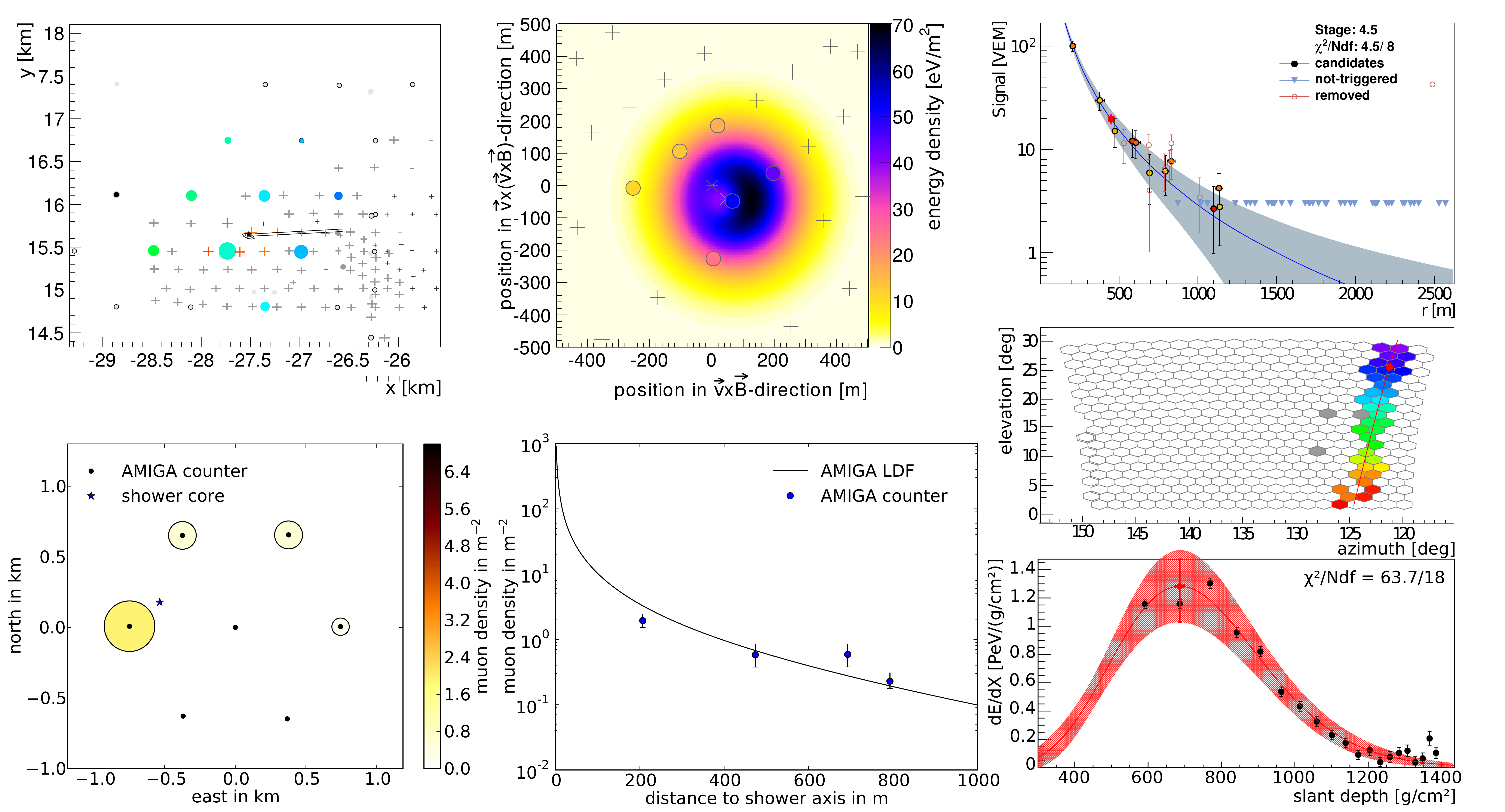}
\caption{Example of an event measured in all four detectors of the Pierre Auger Observatory. Upper left panel: map of AERA and SD stations with signal. Colours correspond to timing, the size of the circles and crosses to the signal strength. The black lines indicate the arrival direction reconstructed from AERA and SD. Upper middle panel: radio 2D-LDF. Upper right panel: SD LDF. Lower left panel: map of muon detector stations with signal. The size and colour corresponds to the muon density. Lower middle panel: muon LDF. Middle right panel: shower trace in FD camera. Lower right panel: Longitudinal profile of the shower measured by the FD. }
\label{fig:hybridEvent}
\end{figure}

The ratio between the number of electrons and muons in the shower is correlated to the mass of the primary particle.
For heavier nuclei, the shower development is faster, i.e. it takes less interaction generations in the cascade until the shower maximum.
This leads to a smaller ratio of numbers of electrons and muons at the shower maximum.
Hence, measuring the electrons and muons separately yields information about the primary particle type.\par
Due to featuring different detector types at one site, the enhancement area of the Pierre Auger Observatory is a perfect place to prove the principle of complementary detection with AERA as a radio detector for the electromagnetic component and AMIGA as a muon counting detector.
While particle detectors only see a snapshot of the shower of the moment it arrives at the detector, AERA does a calorimetric measurement of the electromagnetic part of the shower.
This gives the number of electrons in the whole shower and especially around the shower maximum.
The results can be cross-checked and calibrated with X$_{\textrm{max}}$ measurements of AERA as well as of FD.
In addition, a combined measurement with the X$_{\textrm{max}}$ of AERA around the clock leads to a higher accuracy in the mass sensitivity through improved statistics.
After around two years of combined data taking, more than 200 hybrid detected events are available for analysis.
In figure \ref{fig:hybridEvent}, an example of a multi-hybrid event is shown, which was measured by all four detectors.\par
Another approach aims at inclined showers.
The more inclined the shower, the larger the footprint of the radio emission on ground.
Therefore, the detection efficiency for a radio-antenna field like AERA increases with the zenith angle \cite{inclined}.
The shower itself is older for larger zenith angles when it reaches ground and the electromagnetic part has mainly died out.
Only the muons reach ground and are measured by particle detectors like the water-Cherenkov stations of the Pierre Auger Observatory.
Hence, the electrons and muons can be measured separately by a combination of radio antennas and particle detectors, which enables measurements of the cosmic ray composition even for inclined showers.

\section{Conclusion}
AERA detects several thousand cosmic ray events per year.
Polarization measurements have shown that the emission mechanisms are well understood and that the contribution of the Askaryan effect at the AERA site constitutes 14\,\%.
AERA is sensitive to all cosmic ray properties: the arrival direction, energy and mass.
The cosmic ray energy is derived from the energy stored in the radio emission.
It is calibrated against the energy reconstructed with the particle detector and shows a resolution of 17\,\%.
The mass of the primary particle is correlated with the shower maximum, which can be reconstructed with different methods currently developed in AERA.\par 
Different types of detectors at the Pierre Auger Observatory allow for complementary hybrid measurements of air showers.
This enables a separate measurement of the electromagnetic component with AERA and the muonic component with AMIGA, whose ratio contains information about the primary mass.
In addition, for more inclined air showers, AERA has a higher detection efficiency and SD measures purely the muonic component.
Thus, the complementary measurements of AERA and SD give potential for a significantly increased accuracy of the derived cosmic ray composition.

\end{document}